\documentclass[aps,10pt,prl,twocolumn,showpacs]{revtex4-1}
\usepackage{amsmath,amssymb,amsfonts,amsthm}
\usepackage[utf8]{inputenc}
\usepackage{graphicx}
\usepackage[caption=false]{subfig}
\usepackage{bbm}
\usepackage[pdftex,bookmarks=false,colorlinks=true,linkcolor=blue,
citecolor=blue,filecolor=black,urlcolor=blue]{hyperref}

\begin{document}

\title{Dynamic Correlators of Fermi-Pasta-Ulam Chains and Nonlinear Fluctuating Hydrodynamics}


\author{Christian B. Mendl$^{\mathrm{a}}$}
\email{mendl@ma.tum.de}

\author{Herbert Spohn$^{\mathrm{a,b}}$}
\email{spohn@ma.tum.de}
\affiliation{$^{\mathrm{a}}$Zentrum Mathematik,
Technische Universit\"at M\"unchen,
85747 Garching,
Germany}
\affiliation{$^{\mathrm{b}}$Physik Department,
Technische Universit\"at M\"unchen,
85747 Garching,
Germany}

\date{\today}

\begin{abstract}
We study the equilibrium time correlations for the conserved fields of classical anharmonic chains 
and argue that their dynamic correlator can be predicted on the basis of nonlinear fluctuating hydrodynamics. In fact our scheme is 
 more general and would cover also other one-dimensional hamiltonian systems, for example classical and quantum fluids. 
 Fluctuating hydrodynamics is a nonlinear system of conservation laws with noise. For a single mode it is equivalent to the noisy Burgers equation, for which explicit solutions are available. Our focus is the case of several modes. No exact solutions have been found so far and we rely on 
 a one-loop approximation.
 The resulting mode-coupling equations have a quadratic memory kernel  and describe the time evolving $3\times3$ correlator matrix
 of the locally conserved fields. Long time asymptotics is computed analytically and finite time properties are obtained through a numerical simulation of the mode-coupling equations. 
 \end{abstract}

\maketitle

To predict the dynamic correlator of anharmonic chains is still a theoretical challenge. In higher dimensions 
fluctuating hydrodynamics serves as a convenient starting point~\cite{Forster,RdeL,LL}. But, as recognized already in the 1970ies~\cite{Ernst,FNS}, in one dimension, while the static correlations are of short range, the dynamic current-current correlations generically have an anomalously slow decay. In particular the transport coefficients, required as an input for fluctuating hydrodynamics, are divergent formal expressions. 
There have been huge efforts, both through theoretical investigations and numerical simulations, to more precisely characterize this anomalous behavior (~\cite{Livi,LiviL,LD1,HZ1,LD2,HZ2} for a partial list on  FPU chains only, 
a detailed and commented discussion of the literature can be found in Sect. 7 of~\cite{Spo}). Here we argue that, in one dimension, linear fluctuating hydrodynamics has to be extended to a \textit{nonlinear} version, which will be outlined below. 
Compared to related contributions~\cite{LiLe,vB}, our main advance is to treat the full system of coupled conserved modes and to run time-dependent numerical simulations of the respective mode-coupling equations. In our simulations we use effective coupling constants, which are computed exactly for the particular anharmonic chain under consideration.
Thereby time-resolved predictions are provided which can be tested against molecular dynamics. 

For anharmonic chains, the locally conserved fields are elongation, momentum, and energy. 
Fluctuating hydrodynamics provides a mesoscopic description  of the dynamics of these fields. To illustrate the general framework it is instructive to first recall the simpler case of a single conserved field,
here denoted by  $\tilde{u}(x,t)$, space $x \in \mathbb{R}$,
time $t$. On the macroscopic scale it satisfies the conservation law
\begin{equation}
\label{eq:conslaw1D}
\partial_t \tilde{u}(x,t) + \partial_x j(\tilde{u}(x,t)) = 0
\end{equation}
with given current function $j(\tilde{u})$. We want to study the fluctuations relative to a uniform background $\mathfrak{u}$,
i.e. $\tilde{u}(x,t) = \mathfrak{u} + u(x,t)$, hence expand~\eqref{eq:conslaw1D} to \textit{second} order in $u$ and add dissipation plus noise, resulting in the Langevin equation
\begin{equation}
\label{eq:noisyBurgers}
\partial_t u +\partial_x\big(j'(\mathfrak{u}) u + \tfrac{1}{2} j''(\mathfrak{u}) u^2 - D\partial_x u + \xi\big) = 0\,,
\end{equation} 
where $\xi(x,t)$ is space-time white noise of strength $\sigma$. Since $u$ models the deviations from uniformity, we consider the mean zero, space-time stationary process $u(x,t)$ governed by 
~\eqref{eq:noisyBurgers}. Then, at fixed time $t$, the spatial statistics  is white noise,
$\langle u(x,t) u(x',t) \rangle = \chi \delta(x-x')$, $\chi = \sigma/2D$, which reflects that the static correlations 
of an underlying microscopic model decay exponentially fast. Of particular interest is the
correlator $S(x,t) = \langle u(x,t) u(0,0)\rangle$, $S(x,0) = \chi \delta(x)$. Its large scale behavior will be dominated by the nonlinearity,
but dissipation and noise are required to maintain the proper steady state. Eq.~\eqref{eq:noisyBurgers} is the noisy Burgers equation, equivalently the spatial derivative of the one-dimensional KPZ equation~\cite{KPZ}. There is an exact computation of $S(x,t)$ using replica~\cite{IS}. In particular one knows the universal long time limit,
\begin{equation}
\label{eq:limitKPZ}
S(x,t) = \chi ( \lambda_\mathrm{B} |t|)^{-2/3}f_{\mathrm{KPZ}}\big((\lambda_\mathrm{B}|t|)^{-2/3}(x - j'(\mathfrak{u})t)\big)\,,
\end{equation}
valid for large $x,t$ with $\lambda_\mathrm{B} = \sqrt{2\chi}\, | j''(\mathfrak{u})|$. 
Because of the nonlinearity the spreading is faster than diffusive. 
Note that  $D,\sigma$
appear in Eq.~\eqref{eq:limitKPZ} only through the static susceptibility $\chi$. Identical scaling properties have 
been derived also for stochastic lattice gases~\cite{PSp,FSp}.
 The universal scaling function $f_{\mathrm{KPZ}}$ can be written in terms of a Fredholm determinant 
and has been computed with great precision~\cite{PSp1}.  Interpreting the $u$-field as the slope of a moving front, Eq.~\eqref{eq:limitKPZ} and related predictions  have been confirmed for growth processes in the plane,
both in experiments  on slow combustions fronts
~\cite{MMMT05} and on turbulent liquid crystals~\cite{TS} and numerically through Monte Carlo simulations of Eden cluster growth~\cite{Braz}.

To handle anharmonic chains, we have to extend the above scheme to several components.  We use $\alpha$ as component index.
Then Eq.~\eqref{eq:conslaw1D} generalizes to
\begin{equation}\
\label{eq:conslawcomp}
\partial_t \tilde{u}_\alpha +\partial_x j_\alpha(\vec{\tilde{u}}) = 0\,,\quad \alpha = 1,\dots,n\,,
\end{equation}
$\vec{u} = (u_1,\dots,u_n)$. Expanding as $\tilde{u}_\alpha = \mathfrak{u}_\alpha + u_\alpha$, the coefficients of the linearized equation are
\begin{equation}
A_{\alpha\beta}(\vec{\mathfrak{u}}) = \frac{ \partial j _{\alpha}(\vec{\mathfrak{u}})}{\partial u_{\beta}}
\end{equation}
and the coefficients of the quadratic part are given by the Hessians
\begin{equation}
H^{\alpha}_{\beta\gamma}(\vec{\mathfrak{u}}) = \frac{\partial^2 j_\alpha(\vec{\mathfrak{u}})}{\partial{u_\beta}\partial{u_\gamma}}\,.
\end{equation}
Since the background $\vec{\mathfrak{u}}$ is already prescribed, we will suppress it in our notation. For anharmonic chains
the static correlations of the conserved fields are $\delta$-correlated in space. More generally, rapid decay of static correlations is assumed and hence, at fixed $t$, $u_{\alpha}(x,t)$  is modeled as white noise with covariance
$\langle u_{\alpha}(x,t) u_\beta (x',t) \rangle = C_{\alpha\beta} \,\delta(x-x')$, where $C$ is the $n\times n$ susceptibility matrix with $ C_{\alpha\beta} = C_{\beta\alpha} $. 
As discussed in~\cite{FSS}, $C$ and $A$ are related through
\begin{equation}
AC = CA^{\mathrm{T}}\,,
\end{equation}
transpose denoted by $^\mathrm{T}$, which implies that $A$ has real eigenvalues. We use the eigenvectors of $A$ to construct a linear transformation in component space such that the $\alpha$-th component of the new field travels with a definite velocity,
say $c_\alpha$, at the linearized level. The transformed field components are commonly called normal modes,
denoted here by $\vec{\phi} $. Then $\vec{\phi} = R\vec{u}$ with the $n\times n$ matrix $R$ acting in component space
only.
In addition we require that initially the normal modes are statistically uncorrelated. 
Hence $R$ has to satisfy $RAR^{-1} = \mathrm{diag}(c_1,\dots,c_n)$ and $RCR^\mathrm{T} = 1$, where  both properties together determine $R$ uniquely up to an overall sign.

We now expand in Eq.~\eqref{eq:conslawcomp} to second order in $\vec{u}$, transform to normal modes, and add dissipation plus noise, resulting in the statistical field theory
\begin{equation}
\label{eq:normalmodesPDE}
\partial_t \phi_\alpha + \partial_x\big(c_\alpha \phi_\alpha + \langle \vec{\phi}|G^{\alpha} \vec{\phi}\rangle  - \partial _x (D\vec{\phi})_\alpha
 + \xi_\alpha\big) = 0\,,
\end{equation}
$\alpha = 1,\dots,n$, where 
\begin{equation}\label{27}
G^\alpha = \sum_{\alpha' = 1}^n \tfrac{1}{2} R_{\alpha\alpha'} (R^{-1})^\mathrm{T}H^{\alpha'}R^{-1}\,.
\end{equation}
The diffusion matrix $D$ is positive definite. $\xi_\alpha(x,t)$ is space-time white noise with strength
\begin{equation}
\label{eq:xi_noise}
\langle\xi_\alpha(x,t) \xi_\beta(x',t')\rangle = 2 D_{\alpha\beta}\delta(x - x')\delta(t - t')\,.
\end{equation}
As before, since $\vec{\phi}$ models the deviation from uniformity, we consider the mean zero, stationary process $\vec{\phi}(x,t)$ governed by~\eqref{eq:normalmodesPDE}. In the linear case, $G^\alpha =0$, $\vec{\phi}(x,t)$ is a Gaussian process,  which for fixed $t$ has white noise  statistics with independent components of unit strength, as imposed by $RCR^{\mathrm{T}} = 1$. Note that nonlinear fluctuating hydrodynamics requires as microscopic input only the average currents $j_{\alpha}$, more precisely $A, H^{\alpha}$, and the susceptibility $C$.

Coupled Langevin equations of the form~\eqref{eq:normalmodesPDE} have been proposed and studied before in disguise. Introducing the height $h_\alpha$ through $\partial_x h_\alpha = u_\alpha$, Eq.~\eqref{eq:normalmodesPDE}
turns into the coupled KPZ equations in one dimension, which describe dynamic roughening of directed lines~\cite{EK92}, sedimenting colloidal suspensions~\cite{RS97,LRFB98}, stochastic lattice gases~\cite{DBBR01,FSS}, and magnetohydrodynamics~\cite{Yan97,FD98}. The application to one-dimensional Hamiltonian systems is novel, however.

Equipped with the above frame, let us turn to anharmonic chains, for which purpose 
we first have to figure out the conserved fields and their macroscopic Euler equations. The chain consists of $N$ particles, position $q_j$, momentum $p_j$, $j = 1,\dots,N$, unit mass, and is governed by the Hamiltonian
\begin{equation}
H = \sum_{j=1}^N\big( \tfrac{1}{2} p_j^2 + V(q_{j+1} - q_j)\big)\,,
\end{equation}
where periodic boundary conditions of the form $q_{N+1} = q_1 + L$ are imposed. A prototypical potential is the FPU choice
$V(y) = \tfrac{1}{2}y^2 + \tfrac{1}{3}\mathsf{a}y^3+ \tfrac{1}{4}y^4$. The locally conserved microscopic fields 
are elongation $r_j = q_{j+1} - q_j$, momentum $p_j$, and energy $e_j = \tfrac{1}{2}p_j^2 + V(r_j) 
$. Following our blueprint we collect them as the three-vector $\vec{g}$ with $g_1(j,t)= r_j(t)$,
$g_2(j,t) = p_j(t)$, and
$g_3(j,t) = e_j(t)$.
In a microcanonical simulation one fixes the elongation per particle, $\ell$, as $L = N\ell$,
the momentum per particle, $\mathsf{u}$, as $\sum_{j=1} ^N p_j = N \mathsf{u}$, and the energy per particle,
$\mathfrak{e}$, as $H = N\mathfrak{e}$. Computationally, it is convenient to switch to the canonical pressure ensemble. Then $\ell$ is conjugate to the pressure $p$ and the internal energy $\mathsf{e}$ to the inverse temperature $\beta$.
In the canonical ensemble, $\{r_j,p_j\}$ are independent random variables. The distribution of $p_j$ is a Maxwellian shifted by $\mathsf{u}$, and the distribution of $r_j$ is given by $Z^{-1}\exp[-\beta(V(y) +py)] = \langle \cdot \rangle_{p,\beta}$ with partition function $Z = \int d y \exp[-\beta(V(y) +py)]$. Clearly the pressure equals the average force
acting on a specified particle. The microcanonical and canonical parameters are related through
\begin{equation}
\ell = \langle y\rangle_{p,\beta}\,,\quad \mathsf{e} = \frac{1}{2\beta} + \langle V(y)\rangle_{p,\beta}\,.
\end{equation}

On the hydrodynamic scale the average conserved fields $\langle g_\alpha(j,t)\rangle$ are slowly varying and approximated by the continuum fields $\tilde{u}_\alpha(x,t)$, where $x$ stands for the continuum approximation of the particle index $j$. From the microscopic conservation laws together with local equilibrium one deduces
the hydrodynamic currents 
\begin{equation}
j_\ell = -\mathsf{u}\,,\,\, j_{\mathsf{u}} = p(\ell, \mathfrak{e} - \tfrac{1}{2}\mathsf{u}^2)\,,\,\,
j_{\mathfrak{e}} = \mathsf{u} p(\ell, \mathfrak{e} - \tfrac{1}{2}\mathsf{u}^2)\,,
\end{equation}
which, when inserted in Eq.~\eqref{eq:conslawcomp}, result in the Euler hydrodynamics of the anharmonic chain.
\begin{figure}[!ht]
\centering
\includegraphics[width=0.9\columnwidth]{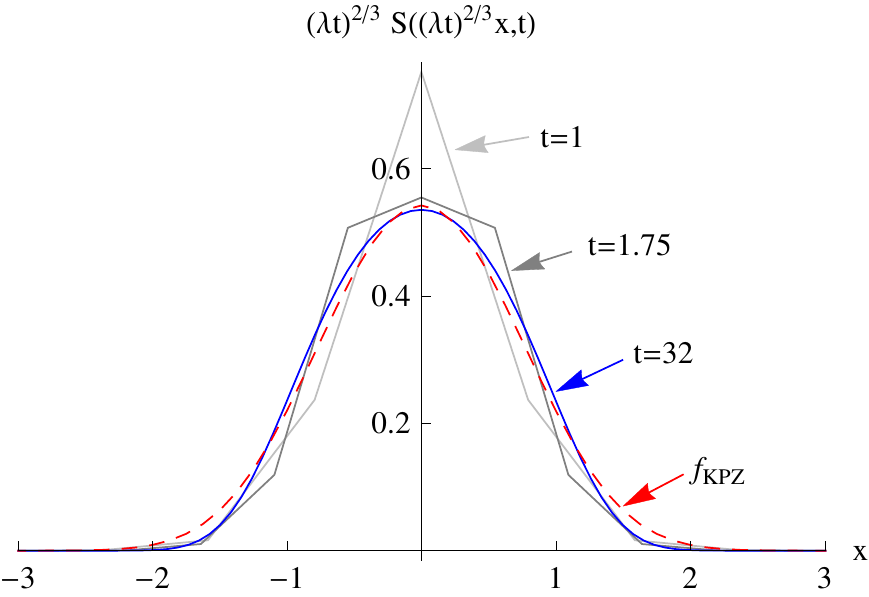}
\caption{(Color online) Time sequence of the numerical mode-coupling solution, $n = 1$, $G^{1}_{11} = \tfrac{1}{2}$, rescaled by $(\sqrt{2}t)^{2/3}$.  The curves are piecewise-linear due to the underlying spatial lattice.
The exact KPZ scaling function is shown in dashed red.}
\label{fig:Sxrescale1D}
\end{figure}

Without loss of generality the equilibrium state of the chain is taken at $\mathsf{u} = 0$. The linear transformation $R$, the velocities $c_\alpha$, and the coupling coefficients $G^\alpha$ are computed in terms of at most third order cumulants involving $y,V(y)$ with average 
$\langle \cdot \rangle_{p,\beta}$. These integrals and the somewhat unwieldy required
substitutions are easily performed using Mathematica. There are three modes: the heat mode, $\alpha = 0$, with velocity $c_0 = 0$ and two sound modes, $\alpha = \pm 1$, with velocity $c_\sigma = \sigma c$, $\sigma = \pm 1$, where $c$ is the sound speed,
\begin{equation}
c^2 = -\partial_\ell p + p\partial_\mathsf{e} p \,.
\end{equation}

The microscopic equilibrium time correlations of the conserved fields in normal mode representation are defined by ($^\sharp$ signals normal mode)
\begin{equation}
\label{eq:22}
S^\sharp_{\alpha\beta}(j,t) = \langle (R\vec{g})_\alpha(j,t) (R\vec{g})_\beta(0,0)\rangle_\mathrm{c} \,,
\end{equation}
the index $\mathrm{c}$ standing for second cumulant. Our central claim is that the normal mode correlations of the chain are approximated for large $x,t$ as
\begin{equation}
\label{eq:23}
S^\sharp_{\alpha\beta}(j,t) \simeq \langle \phi_\alpha(x,t) \phi_\beta(0,0)\rangle  = S^{\sharp\phi}_{\alpha\beta}(x,t)\,.
\end{equation}
This leaves us with the task to work out the correlator $S^{\sharp\phi}$ for the stochastic field theory~\eqref{eq:normalmodesPDE}. With no exact solution at hand, we rely on the mode-coupling equations in one-loop approximation. But before, since the three modes travel with distinct velocities,  they decouple for long times and only the self-interaction proportional to $G^\alpha_{\alpha\alpha}$ will contribute. Since  generically 
 $G^{1}_{11} \neq 0$, and $G^{1}_{11} = - G^{-1}_{-1-1}$, the two sound modes are expected to satisfy the KPZ scaling~\eqref{eq:limitKPZ} with 
 the substitutions $\chi = 1$,
$ j'(\mathfrak{u}) = \sigma c$ and $ \lambda_\mathrm{B} = 2 \sqrt{2} |G^{\sigma}_{\sigma\sigma}|$, $\sigma = \pm 1$. The decoupling of modes is convincingly confirmed in a two-component lattice gas~\cite{FSS}.
For the heat mode our argument fails, since $G^0_{00}= 0$ always. Note that for the popular case of an even potential,
$V(y) = V(-y)$, at $p=0$ also $G^1_{11} = 0$ implying that all three modes are non-KPZ.

The derivation of the mode-coupling equations is explained in~\cite{Spo}. To be concise we only display 
the diagonal approximation, for which $S^{\sharp}_{\alpha\beta}(x,t) \simeq \delta_{\alpha\beta}f_{\alpha}(x,t)$ is assumed. Switching to Fourier space and adopting the standard conventions for discrete Fourier transforms, the mode-coupling equations then simplify to
\begin{multline}
\label{eq:mcfourier}
\partial_t \hat{f}_{\alpha}(k,t) = -\mathrm{i}\, c_{\alpha} \sin(2\pi k) \hat{f}_{\alpha}(k,t) 
- 2 (1 - \cos(2\pi k))\\
\times\Big(D_{\alpha} \hat{f}_{\alpha}(k,t) 
+ \int_0^t d s \, \hat{M}_{\alpha\alpha}(k,s) \, \hat{f}_{\alpha}(k,t-s)\Big)\,,
\end{multline}
$ \alpha = 1,\dots,n$, with memory kernel 
\begin{equation}
\label{eq:memkernelDiag}
\hat{M}_{\alpha\alpha} (k,t) = 2\sum^n_{\beta,\gamma=1}(G^{\alpha}_{\beta\gamma})^2 
 \int_{-1/2}^{1/2} d q \, \hat{f}_{\beta}(k-q,t) \hat{f}_{\gamma}(q,t)\,.\end{equation}
Numerically we always simulate the dynamics of the full correlator matrix. For a wide range of parameters,
after some transient time the off-diagonal matrix elements decay and are always  by an order of magnitude smaller than the diagonal ones.

The special case $n=1$ is discussed already in~\cite{vBKS}, see~\cite{CM} for a first numerical integration.
\begin{figure*}[!ht]
\centering
\subfloat[$t = 12$]{
\includegraphics[width=0.3\textwidth]{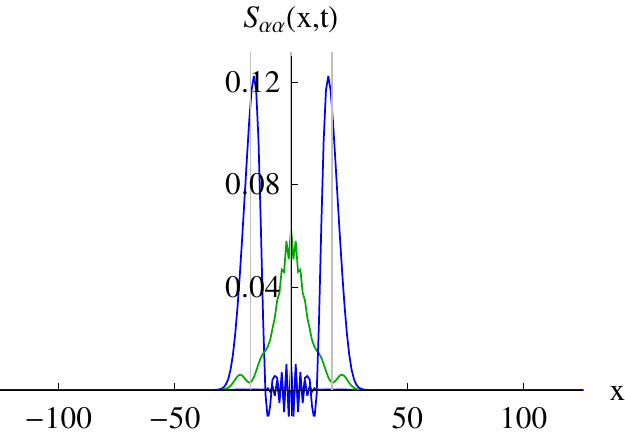}}
\subfloat[$t = 32$]{
\includegraphics[width=0.3\textwidth]{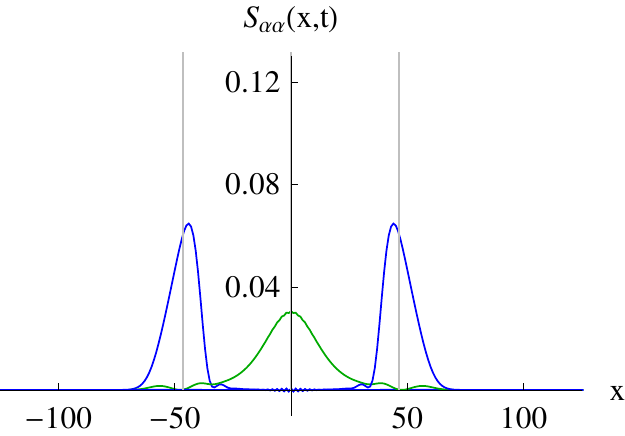}}
\subfloat[$t = 64$]{
\includegraphics[width=0.3\textwidth]{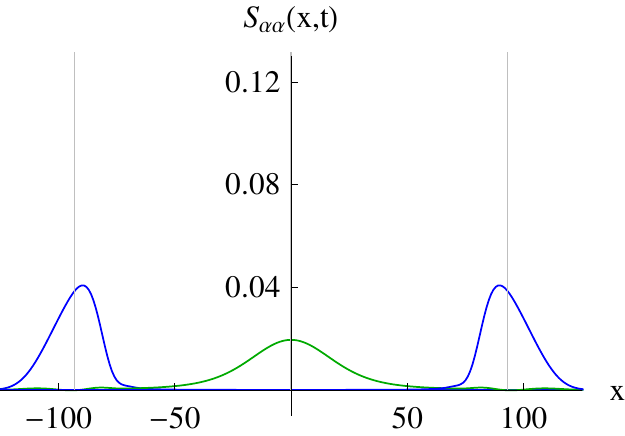}}\\
\subfloat[heat mode peak]{
\includegraphics[width=0.4\textwidth]{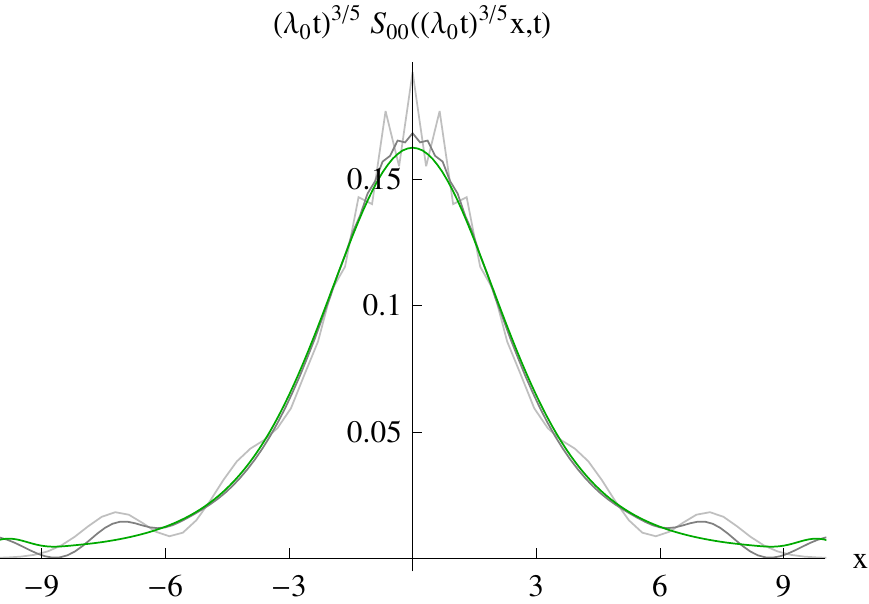}}
\hspace{0.05\textwidth}
\subfloat[right sound mode peak]{
\includegraphics[width=0.4\textwidth]{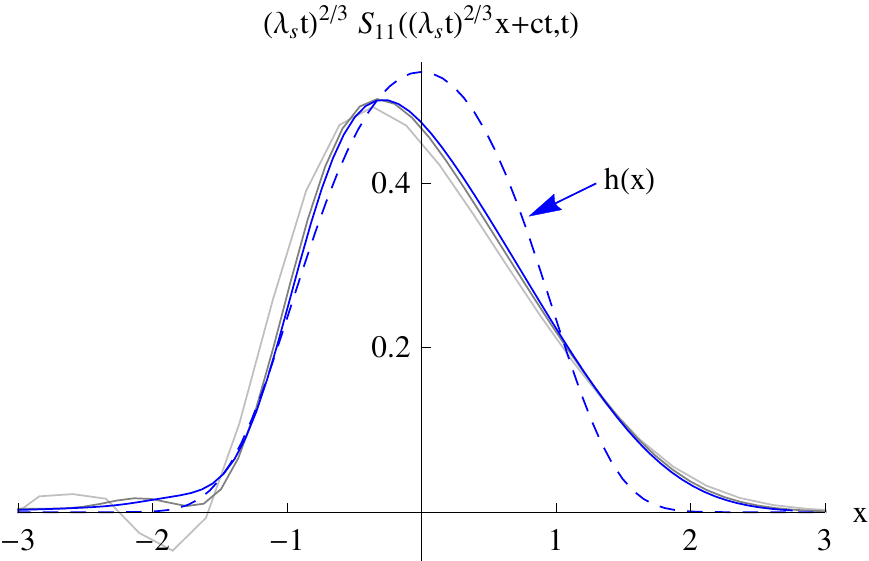}}
\caption{(Color online) Time sequence of normal mode correlations for the FPU chain with $\mathsf{a} = 2$, $p = 1$, $\beta = 2$, and $\mathsf{u} = 0$ (a), (b), (c). Magnification of the central heat mode peak (d) and the right sound mode peak (e) in suitably rescaled coordinates.}
\label{fig:Sstandard3D}
\end{figure*}
In Fig.~\ref{fig:Sxrescale1D} we display a  time sequence for a single mode with $G^{1}_{11} = \tfrac{1}{2}$.  For $t \gtrsim 32$ the scaled solution remains stationary.  The asymptotic scaling function differs from $f_{\mathrm{KPZ}}$ by a few percent only. On this basis
we expect that such a precision extends to several modes.

As a representative example for anharmonic chains we choose the FPU potential with $\mathsf{a} = 2$ and $\mathsf{u} = 0$, 
$p=1$, $\beta = 2$, resulting in $c = 1.455$, which  are commonly used parameters in molecular dynamics simulations. $V(y) + py$ has a single minimum at $y = -1.755$. We stress that simulations can be performed for any choice of the potential and thermodynamic parameters at minimal numerical efforts. With the theoretically determined velocities and couplings, the mode-coupling equations are iterated in time, using Fourier space representation as in~\eqref{eq:mcfourier}, in such a way that the values of the memory kernel $\hat{M}_{\alpha\alpha'}(k,s)$ for $s < t$ can be stored and re-used.  The time and momentum variables are discretized by a uniform grid. In
Fig.~\ref{fig:Sstandard3D} the grey vertical lines at $ \pm ct$ indicate the predicted position of the sound mode peaks. The off-diagonal elements of $S^{\sharp\phi}$ are essentially zero. In the time sequence we display the superimposed diagonal normal mode
correlations (area $1$ under each curve). More details are provided in the blow-up. For the heat mode peak one observes oscillations which move away from the center and eventually die out. The tail of the heat mode peak is cut at the location of the sound mode. At the longest available time the sound mode peaks are still asymmetric and have not yet reached their asymptotic shape.

Theoretically the scaling function for the heat mode peak is obtained by inserting  the known asymptotic form 
of $\hat{f}_{\pm1}$ in~\eqref{eq:memkernelDiag} with $\alpha = 0$. Solving the then linear memory equation~\eqref{eq:mcfourier} results in the symmetric Levy $5/3$ distribution, $\hat{f}_0(k,t) = \exp[-|k|^{5/3}
\lambda_0 |t|]$ with computed non-universal coefficient $\lambda_0$. The Fourier transform of $\hat{f}_0(k,t)$
is plotted as green curve in Fig.~\ref{fig:Sstandard3D}~(d).
The Levy $5/3$ distribution for the heat mode peak has been guessed earlier based on the molecular dynamics of the hard-point gas with alternating masses \cite{Cipriani2005}, reconfirmed in case a maximal distance between the hard-point particles is imposed~\cite{HardPoint2007}, and also for FPU chains~\cite{Denisov2011}. Currently the Levy distribution is the strongest numerical support of nonlinear fluctuating hydrodynamics. Note that no signal propagates outside the sound cone. 

Based on these and further simulations of the mode-coupling equations for anharmonic chains, the following qualitative picture for the motion of the normal mode peaks in index number space emerges. The sound mode peaks ``rapidly'' decay to a shape function which is centered at $\sigma c t$ and  varies on the scale $t^{2/3}$. The shape function itself is still slowly varying. The couplings $G^0_{\sigma\sigma}$ determine the scaling of the heat mode peak. Since only the integral over the square of the shape function is involved, the heat mode peak rapidly achieves its asymptotic shape in the range $\{|x| \leq ct\}$ with a still slowly varying non-universal constant.
The intermediate time motion of the sound mode peaks is dominated by $G^\sigma_{00}$ and $G^\sigma_{-\sigma-\sigma}$. Assuming already the validity of overall scaling picture, the size of these finite time corrections is estimated to be of the order $t^{-1/15}$, resp. $t^{-1/9}$,
relative to the leading term
which signals that $f_{\mathrm{KPZ}}$ is approached rather slowly. Of course, only a qualitative guideline is presented.
For the precise dynamics all velocities and couplings have to be taken into account.
\noindent\medskip\\
\textit{Conclusions}. We developed a nonlinear extension of fluctuating hydrodynamics applicable to one-dimensional systems, in principle including classical fluids, quantum fluids~\cite{Andreev, Zwerger, L,ArBo13}, and quantum spin chains. Already at the level of the one-loop approximation it is crucial to maintain the couplings between all conserved modes. As applied to anharmonic chains, the numerical solutions of the 
mode-coupling equations provide a realistic picture of the correlation dynamics and, on the limited space-time scale simulated, are consistent with the analytical computations and also with molecular dynamics, as far as available. 

We thank Henk van Beijeren, Patrik Ferrari, Tomohiro Sasamoto, and Hong Zhao
for most useful comments. The research is supported by DFG project SP 181/29-1.

%\bibliographystyle{apsrev}
%\bibliography{references}

\end{document}